\documentclass{raa}
\usepackage{graphicx,times}
\usepackage{natbib}
\usepackage{amssymb,amsmath}
\bibpunct{(}{)}{;}{a}{}{,}
\usepackage{lineno}

\usepackage[a4paper=true,pagebackref=true]{hyperref}
\hypersetup{pdftitle = The title of my PDF,
           pdfauthor = My name,
           pdfsubject= The subject,
           pdfkeywords = keyword1 keyword2 keyword3}
\hypersetup{colorlinks = true,
             linkcolor = green,
             anchorcolor = red,
             citecolor = blue,
             filecolor = red,
             pagecolor = red,
             urlcolor = red}

\renewcommand{\thefootnote}{\fnsymbol{footnote}}

\begin{document}
\title{DmpIRFs and DmpST: DAMPE Instrument Response Functions and Science Tools for Gamma-Ray Data Analysis}
\author{Kai-Kai Duan\inst{1, 2} \footnote{duankk@pmo.ac.cn}, Wei Jiang\inst{1, 3}, Yun-Feng Liang\inst{1} \footnote{liangyf@pmo.ac.cn}, Zhao-Qiang Shen\inst{1, 2}, Zun-Lei Xu\inst{1}, 
        Yi-Zhong Fan\inst{1, 3}, Fabio Gargano\inst{4}, Simone Garrappa\inst{5, 6}, Dong-Ya Guo\inst{7}, Shi-Jun Lei\inst{1}, Xiang Li\inst{1} \footnote{xiangli@pmo.ac.cn}, Mario Nicola Mazziotta\inst{4}, 
        Maria Fernanda Munoz Salinas\inst{8}, Meng Su\inst{1, 9}, Valerio Vagelli\inst{5, 6}, Qiang Yuan\inst{1, 3}, Chuan Yue\inst{1}, Stephan Zimmer\inst{8}}
\institute{ Key Laboratory of Dark Matter and Space Astronomy,
            Purple Mountain Observatory,
            Chinese Academy of Sciences,
            Nanjing 210008, China
            \and
            University of Chinese Academy of Sciences,
            Beijing 100049, China
            \and
            School of Astronomy and Space Science,
            University of Science and Technology of China,
            Hefei, Anhui 230026, China
            \and
            Istituto Nazionale di Fisica Nucleare Sezione di Bari,
            I-70125, Bari, Italy
            \and
            Istituto Nazionale di Fisica Nucleare Sezione di Perugia,
            I-06123 Perugia, Italy
            \and
            Dipartimento di Fisica e Geologia, Università degli Studi di Perugia,
            I-06123 Perugia, Italy
            \and
            Institute of High Energy Physics,
            Chinese Academy of Sciences,
            Beijing, 100049, China
            \and
            Department of Nuclear and Particle Physics,
            University of Geneva, CH-1211, Switzerland
            \and
            Department of Physics and Laboratory for Space Research, 
            University of Hong Kong, Pok Fu Lam, Hong Kong, China
}
\abstract{ GeV gamma ray is an important observation target of DArk Matter Particle Explorer (DAMPE)
           for indirect dark matter searching and high energy astrophysics.
           We present in this work a set of accurate instrument response functions of DAMPE (DmpIRFs) including the effective area, point-spread function and energy dispersion that are crucial for the gamma-ray data analysis based on the high statistics simulation data.
           A dedicated software named DmpST is developed to facilitate the scientific analyses of DAMPE gamma-ray data.
           Considering the limited number of photons and the angular resolution of DAMPE,
           the maximum likelihood method is adopted in the DmpST to better disentangle different source components.
           The basic mathematics and the framework regarding this software are also introduced in this paper.
}
\authorrunning{K.-K. Duan et al.}
\titlerunning{DmpIRFs and DmpST}
\maketitle

%
\section{Introduction}
DArk Matter Particle Explorer (DAMPE) is a high energy cosmic-ray and gamma-ray observatory
(\citealt{Chang2014}, \citealt{CHANG20176}).
It contains four sub-detectors: a Plastic Scintillation Detector (PSD),
a Sillicon-Tungsten tracKer-converter (STK), a BGO calorimeter (BGO) and a NeUtron Detector (NUD).
The PSD that measures the charge of particles also acts as anti-coincidence detector for gamma-ray observation.
The STK measures the trajectories of charged particles, as well as the photons that are converted into e$^{+}$e$^{-}$ pairs.
The BGO calorimeter measures the energies of incidence particles
and is also able to distinguish the electron and hadron efficiently.
The NUD provides an independent measurement and further improvement for the electron/hadron identification.
The on-orbit calibration have adopted for DAMPE and is expected to operate stably during the next few years (\citealt{AMBROSI201918}, \citealt{Ma2018brb}, \citealt{2018arXiv181009901D}, \citealt{jiangwei}).

Based on the photon selection algorithm described in \citet{Xu2018},
valuable gamma-ray data have been accumulated.
Further scientific analysis of high-level gamma-ray data, however, 
requires detailed knowledge about the instrument response functions (IRFs) of DAMPE, 
i.e., the effective area, the point-spread function (PSF) and the energy dispersion function.
Using the high-statistics simulation data, we have constructed the IRFs for DAMPE gamma-ray observation
in the energy range from 1 GeV to 10 TeV and with the incidence angle from 0$^{\circ}$ to 60$^{\circ}$.

Limited by the relatively low statistics of DAMPE gamma-ray data,
the chi square method is not suitable for the data analysis,
maximum likelihood method (\citealt{1996ApJ...461..396M}) is adopted.
Combining the IRFs and the model of gamma-ray sources,
we can calculate the expected photon number recorded by the detector.
The values, and also the uncertainties, of the parameters in the gamma-ray source model 
then can be estimated by comparing with the real DAMPE observation using the maximum likelihood method.

The data preparation, the convolution with the IRFs and the parameter inference
are realized for DAMPE data analysis using a dedicated software named DmpST, 
which is also developed to facilitate the scientific analysis. 
In this paper, we introduce both the DAMPE IRFs and the DmpST software.

This paper is structured as the following. We first introduce the IRFs of DAMPE in Section 2.
The {\it observing time} and exposure of DAMPE are then described in Section 3.
In Section 4, we introduce the maximum likelihood method for DAMPE gamma-ray data analysis,
followed by a description of the code structures in Section 5. We summarize this work in Section 6.

\section{Instrument Response Functions}
Instrument response functions (IRFs) are the parameterized representations of the instrument performance.
The DAMPE IRFs can be factorized into three parts (\citealt{2012ApJS..203....4A}).
The effective area, $A_{\mathrm{eff}}(E, \hat{v}, s)$, is the product of the geometrical cross-section area,
the probability of gamma-ray conversion and the efficiency of photon selection for a gamma-ray 
with energy $E$ and direction $\hat{v}$ in the detector reference frame.
The $s$ denotes the trigger type (see below).
The Point-spread function (PSF), $P(\hat{v}'; E, \hat{v}, s)$ and the energy dispersion function, $D(E'; E, \hat{v}, s)$
are the probability distributions of the reconstructed direction $\hat{v}'$ and the reconstructed energy $E'$
for a gamma-ray with energy $E$ and direction $\hat{v}$.

Given the spatial and spectral model of the incidence gamma-ray sources, $F(E, \hat{p})$,
where $\hat{p}$ refers to the celestial directions of the gamma-ray sources,
we can convolve the model with the IRFs to predict the distribution of observed photons:
\begin{eqnarray}
\nonumber r(E', \hat{p}', s) = \int \int \int F(E, \hat{p}) A_{\mathrm{eff}}(E, \hat{v}(t; \hat{p}), s) \\
\times P(\hat{v}'(t; \hat{p}'); E, \hat{v}(t; \hat{p}), s) D(E'; E, \hat{v}(t; \hat{p}), s) \mathrm{d}E \mathrm{d}\Omega \mathrm{d}t,
\end{eqnarray}
where $\hat{p}'$ is the reconstructed celestial directions of the gamma-rays.
The integrals are over the time and energy range of interest and the solid angle in the celestial reference frame.

To evalute the DAMPE IRFs, we perform {\tt Geant4}-based Monte Carlo detector simulation to generate pseudo-photons of DAMPE (MC data hereafter). 
We simulate gamma-rays with uniform distribution of incidence direction, 
that can be used to explore the instrument response across the entire field of view (FoV) of DAMPE.
The MC data are generated with an $E^{-1}$ counts spectrum uniformly in the logarithm energy,
and from a sphere with 6 m$^{2}$ cross-sectional area centered on the detector 
to cover the whole energy range and the whole detector of DAMPE.
The directions of the gamma rays are sampled uniformly in solid angle with downward-going directions,
leading to a semi-isotropic incidence flux of the simulated gamma rays.
Here we ignore the back-entering events, because these events would have to traverse a large amount of material 
and thus presumably lose a lot of energy along their way.
Through the same reconstruction and gamma-ray selection algorithm as the on-orbit data, 
the MC data can describe the response of DAMPE for gamma-ray observation accurately (\citealt{Xu2018}).

DAMPE uses two sets of trigger directives for physics data: the pre-scaled Low Energy Trigger (LET) and the Higt Energy Trigger (HET).
The pre-scale factors of LET are different when the detector is in different geographic latitude (\citealt{CHANG20176}).
When the detector is in the low latitude region ($|\phi_{\rm g}| < 20^{\circ}$), the Lower Energy trigger is pre-scaled with a factor of 8;
and at high latitude region ($|\phi_{\rm g}| > 20^{\circ}$) it is 64 pre-scaled.
The IRFs is also divided into two sub-sets, LET IRFs and HET IRFs.

\subsection{Effective area}
Effective area is a numerical function varying with the energy of gamma-ray photon and its incidence direction in the instrument reference frame.
We binned the MC data according to the event energy, incidence angle and trigger type.
The effective area for each bin centered at $E_{i}$, $\theta_{j}$, $\phi_{k}$  with trigger type $s$ is
\begin{eqnarray}
A_{\mathrm{eff}}(E_{i}, \theta_{j}, \phi_{k}, s) = \frac{N_{i, j, k, s}}{N_{\mathrm{sim}, i, j, k}} A_{\mathrm{sim}},
\end{eqnarray}
where $N_{\mathrm{sim}, i, j, k}$ is the number of photons generated in the simulation in each bin,
and $N_{i, j, k, s}$ is the number of photons that passing the selection algorithm with trigger type $s={\rm LET}$ or ${\rm HET}$. 
The $A_{\mathrm{sim}}$ is the cross-section area of the generated sphere in the simulation.

\begin{figure}
\center
\includegraphics[width=10.0cm]{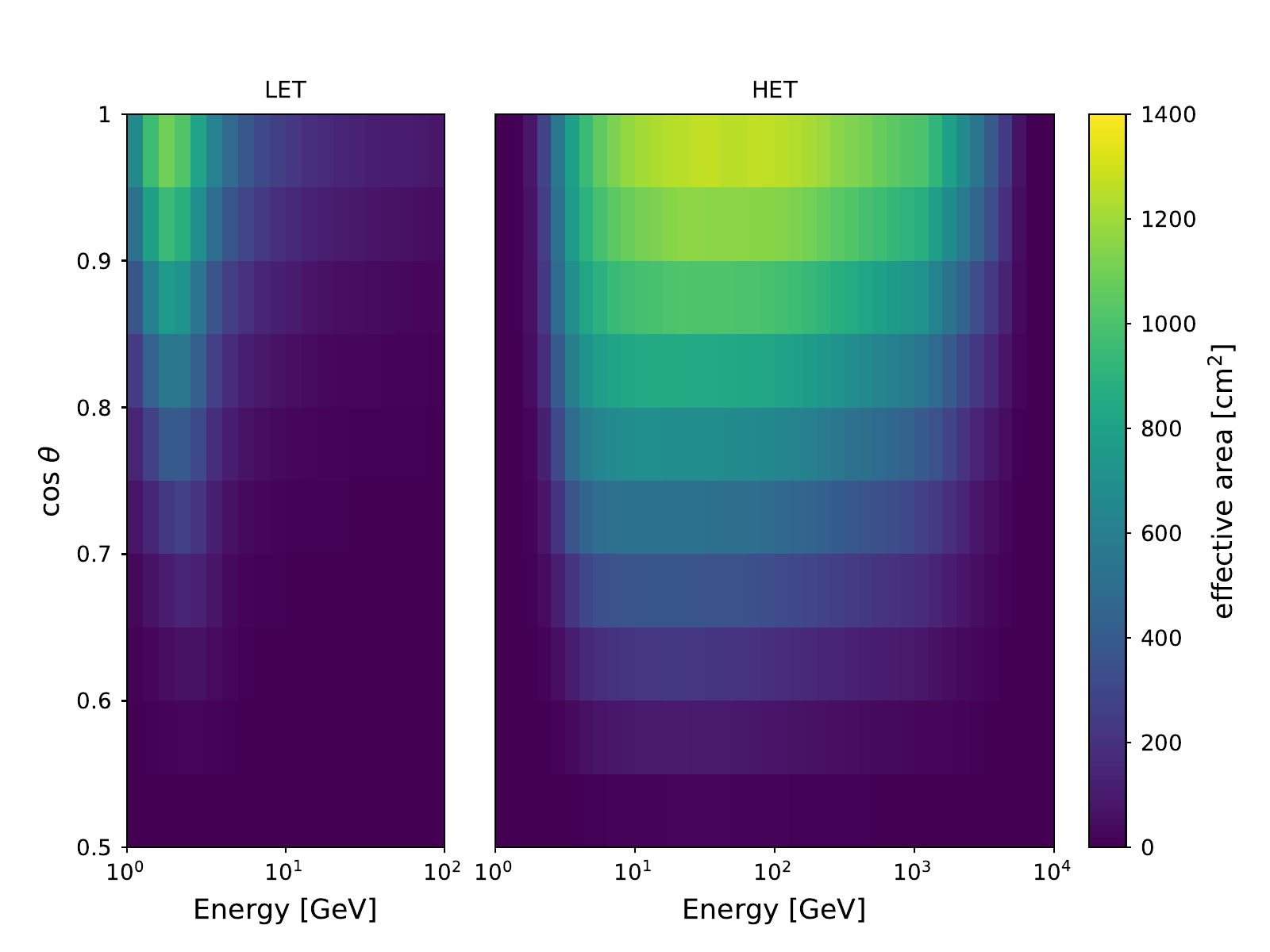}
\caption{The effective area of DAMPE (in units of ${\rm cm}^{2}$) for gamma-ray observation at different energy and incidence direction.
         The energy is in 20 bins from 1 GeV to 100 GeV for LET photons (left panel) and 40 bins to 10 TeV for HET photons (right panel).
         The incidence angle is in 10 bins from 0$^{\circ}$ to 60$^{\circ}$.
         Note that the effective area presented here is averaged over $\phi$.}
\label{AEFF}
\end{figure}

We divide the MC data into 20 energy bins from 1GeV to 100GeV (40 energy bins from 1 GeV to 10 TeV) and 10 angular bins from 0$^{\circ}$ to 60$^{\circ}$ for LET (HET) data. 
Fig. \ref{AEFF} shows the effective area of DAMPE gamma-ray observation as a function of the energy and incidence direction.

\subsection{Point-spread function (PSF)}
The reconstructed direction ($\hat{v}'$) of the photon may deviate from its true value ($\hat{v}$),
and the probability distribution of the deviation $\delta v = |\hat{v}' - \hat{v}|$ is parameterized by the PSF.
The PSF for the DAMPE is related to the inclination angle $\theta$ and the azimuth angle $\phi$ of the incidence photon in the detector reference frame, 
and also the photon's energy and trigger type.
Because the $\phi$ dependence of the PSF is much weaker than the $\theta$ dependence,
we ignore the $\phi$ dependence in the current version of the PSF.

Based on the MC data, we construct a histogram of the angular deviations of the selected gamma-rays 
for each energy and incidence angle bin and for each trigger type.
We find that the form of the Fermi-LAT PSF (\citealt{2012ApJS..203....4A}) can accommodate DAMPE simulation data well.
Accordingly, the PSF histogram is fitted with a double King function,
\begin{eqnarray}
P(x) = f_{\mathrm{core}} K(x_{p}; \sigma_{\mathrm{core}}, \gamma_{\mathrm{core}}) 
       + (1 - f_{\mathrm{core}}) K(x_{p}; \sigma_{\mathrm{tail}}, \gamma_{\mathrm{tail}}),
\end{eqnarray}
where $K(x_{p}; \sigma, \gamma)$ is King function defined as
\begin{eqnarray}
K(x_{p}; \sigma, \gamma) = \frac{1}{2\pi\sigma^{2}} \left( 1-\frac{1}{\gamma} \right)
                        \left[ 1+\frac{1}{2\gamma}\frac{x_{p}^{2}}{\sigma^{2}} \right]^{-\gamma},
\end{eqnarray}
and $x_{p}$ is the scaled angular deviation 

\begin{eqnarray}
x_{p} = \frac{\delta v}{S_{\mathrm{p}}(E, \theta)}.
\end{eqnarray}

The $S_{\mathrm{p}}(E, \theta)$ is the angular resolution (defined as 68$\%$ containment of the angular deviation) at energy $E$ and incidence angle $\theta$.
The functional form of the King profile originates from XMM Newton (\citealt{2004SPIE.5488..103K}, \citealt{2011A&A...534A..34R}) and was later adapted for the Fermi-LAT. 
Note that the King function is normalized, i.e., $\int_{0}^{\infty} 2 \pi x K(x; \sigma, \gamma) \mathrm{d}x = 1$.

\begin{figure}
\center
\includegraphics[width=10.0cm]{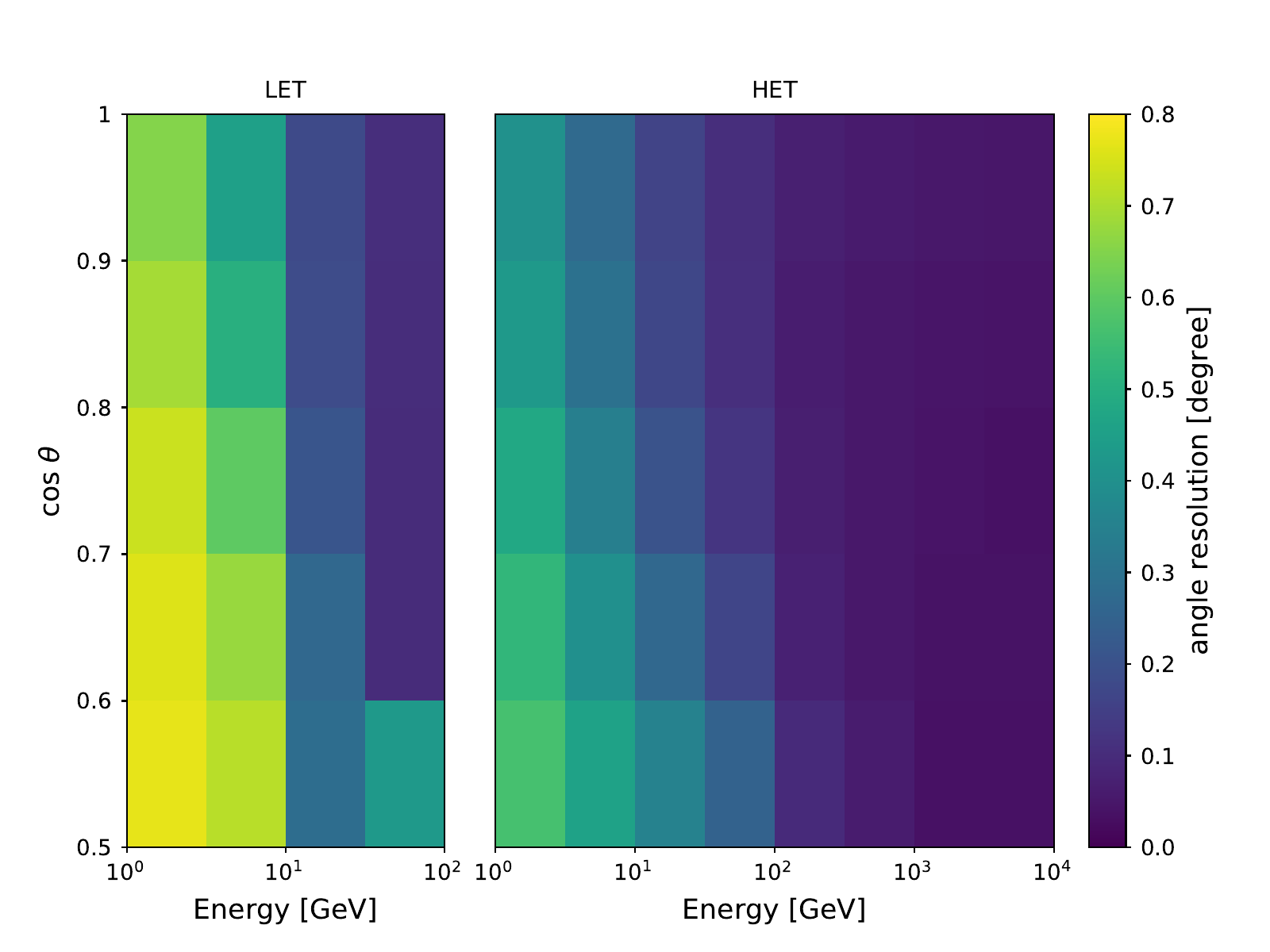}
\caption{The angular resolution of DAMPE (in units of degree) for gamma-ray observation at different energy and incidence direction. 
		 The energy is in 4 bins from 1 GeV to 100GeV for LET photons (left panel) and 8 bins to 10 TeV for HET photons (right panel).
		 The incidence angle is in 5 bins from 0$^{\circ}$ to 60$^{\circ}$.}
\label{PSF}
\end{figure}

\begin{figure}
\center
\includegraphics[width=10.0cm]{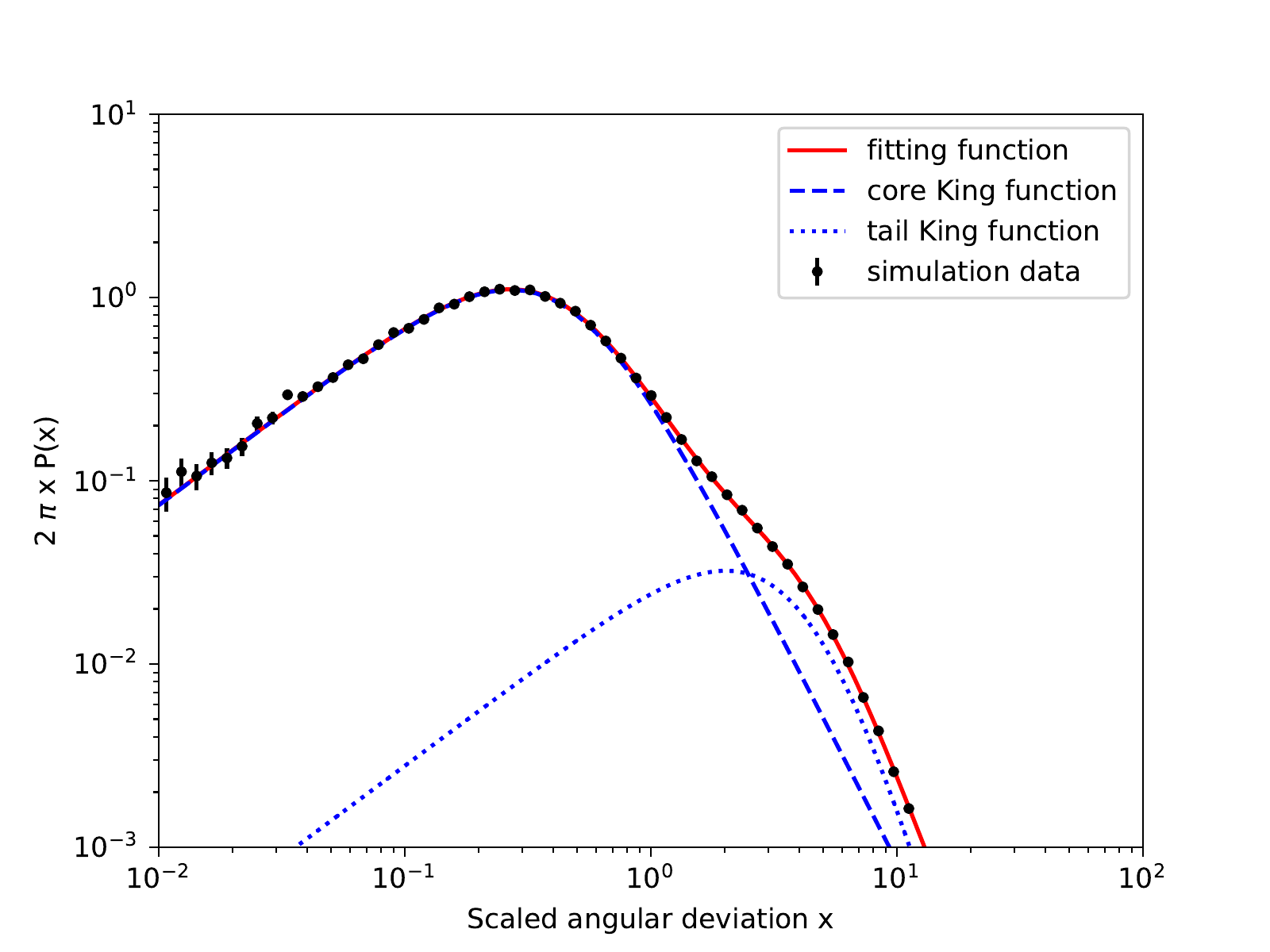} 
\caption{The best fit to the scaled angular deviation with double King function
         in the energy range [3.16, 10] GeV and incidence angle range [25.84$^{\circ}$, 36.87$^{\circ}$] for HET photons.
         The points are the distribution of the scaled angular deviation of the MC data, 
         the dash and dotted line are the core and tail King functions respectively 
         and the solid line is the sum of the two components.
         The reduce $\chi^{2}$ of this fitting is 1.09.}
\label{psf}
\end{figure}

We divide the MC data into 4 energy bins from 1GeV to 100GeV (8 energy bins from 1 GeV to 10 TeV) 
and 5 angular bins from 0$^{\circ}$ to 60$^{\circ}$ for LET (HET) data. 
Fig. \ref{PSF} shows the angular resolution of DAMPE for gamma-ray observation at different energy and incidence direction.
For each bin, the MC data are fitted with above functions and the best-fit parameters of them are derived and stored in the DmpST. 
Fig. \ref{psf} shows an example of the best fit to the scaled angular deviation with the double King function
in the bin of $E\in[3.16, 10]\,{\rm GeV}$ and $\theta\in[25.84^{\circ}, 36.87^{\circ}]$ for HET photons.

\subsection{Energy dispersion}
Energy dispersion function gives the probability of a photon with true energy ($E$) being allocated an energy ($E'$) after the events reconstruction. 
Similar to the PSF, we ignore the $\phi$ dependence and parameterize the energy dispersion as function of scaled energy deviation
\begin{eqnarray}
x_{D} = \frac{E'-E}{S_{\mathrm{D}}(E, \theta)E},
\end{eqnarray}
where the scale $S_{\mathrm{D}}(E, \theta)$ is the energy resolution (defined as the half-width of the 68$\%$ containment range of the energy deviation)
at the bin center of energy $E$ and incidence angle $\theta$.
We fit the MC data with three piecewise functions of the form

\[D(x_{D}) = \begin{cases}
N_{L} R(x_{D}, x_{0}, \sigma_{L}, \gamma_{L}) & \text{if $(x_{D} - x_{0}) < -\bar{x}$} \\
N_{l} R(x_{D}, x_{0}, \sigma_{l}, \gamma_{l}) & \text{if $(x_{D} - x_{0}) \in [-\bar{x}, 0]$} \\
N_{R} R(x_{D}, x_{0}, \sigma_{R}, \gamma_{R}) & \text{if $(x_{D} - x_{0}) > 0 $}\\
\end{cases} \]
\begin{eqnarray}
R(x_{D}, x_{0}, \sigma, \gamma) = N \exp \left( -\frac{1}{2} \left| \frac{x_{D} - x_{0}}{\sigma}\right|^{\gamma} \right).
\end{eqnarray}

\begin{figure}
\center
\includegraphics[width=10.0cm]{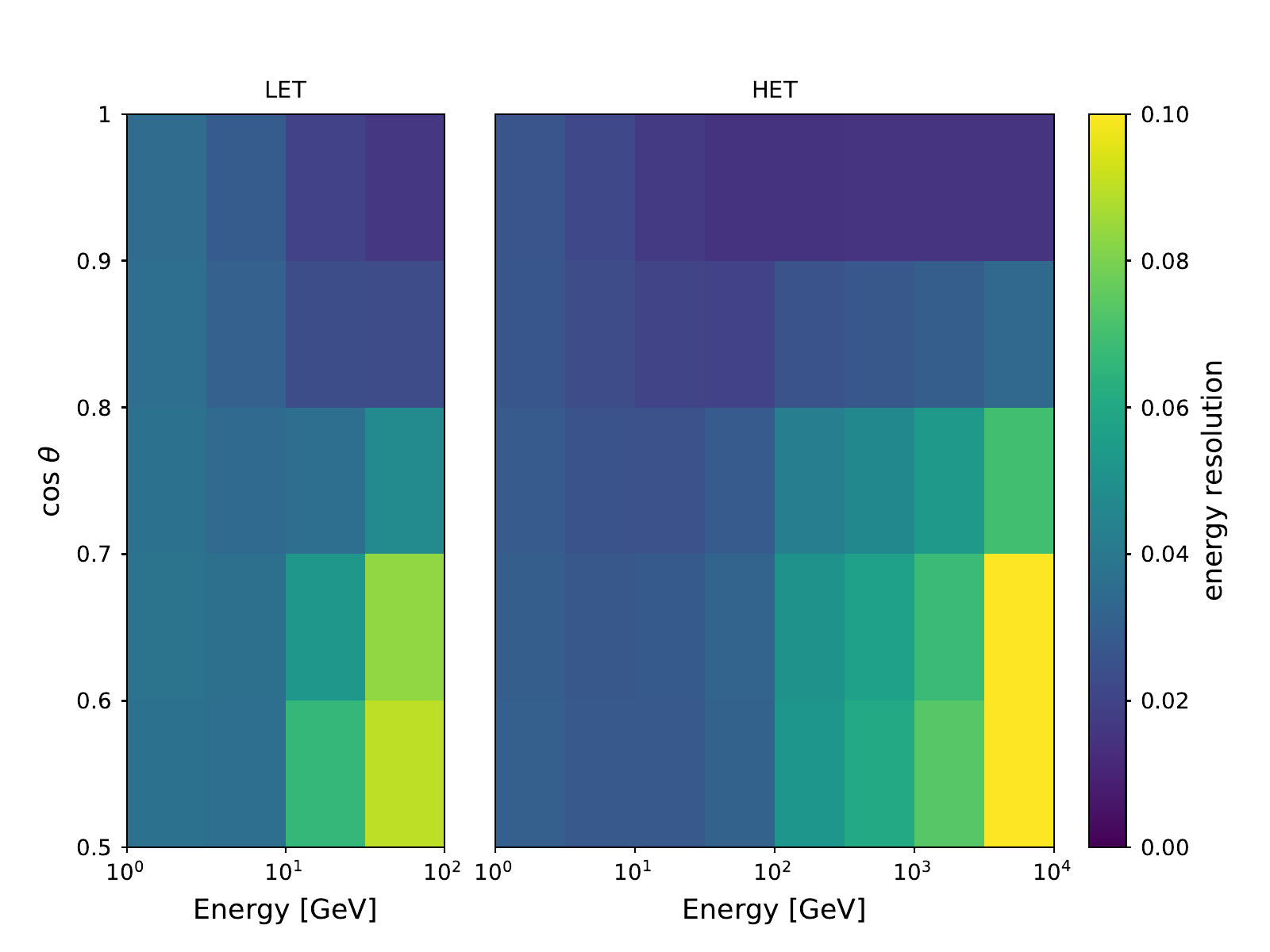}
\caption{The energy resolution of DAMPE (dimensionless) for gamma-ray observation at different energy and incidence direction. 
		 The energy is in 4 bins from 1 GeV to 100GeV for LET photons (left panel) and 8 bins to 10 TeV for HET photons (right panel).
		 The incidence angle is in 5 bins from 0$^{\circ}$ to 60$^{\circ}$.}
\label{EDISP}
\end{figure}

\begin{figure}
\center
\includegraphics[width=10.0cm]{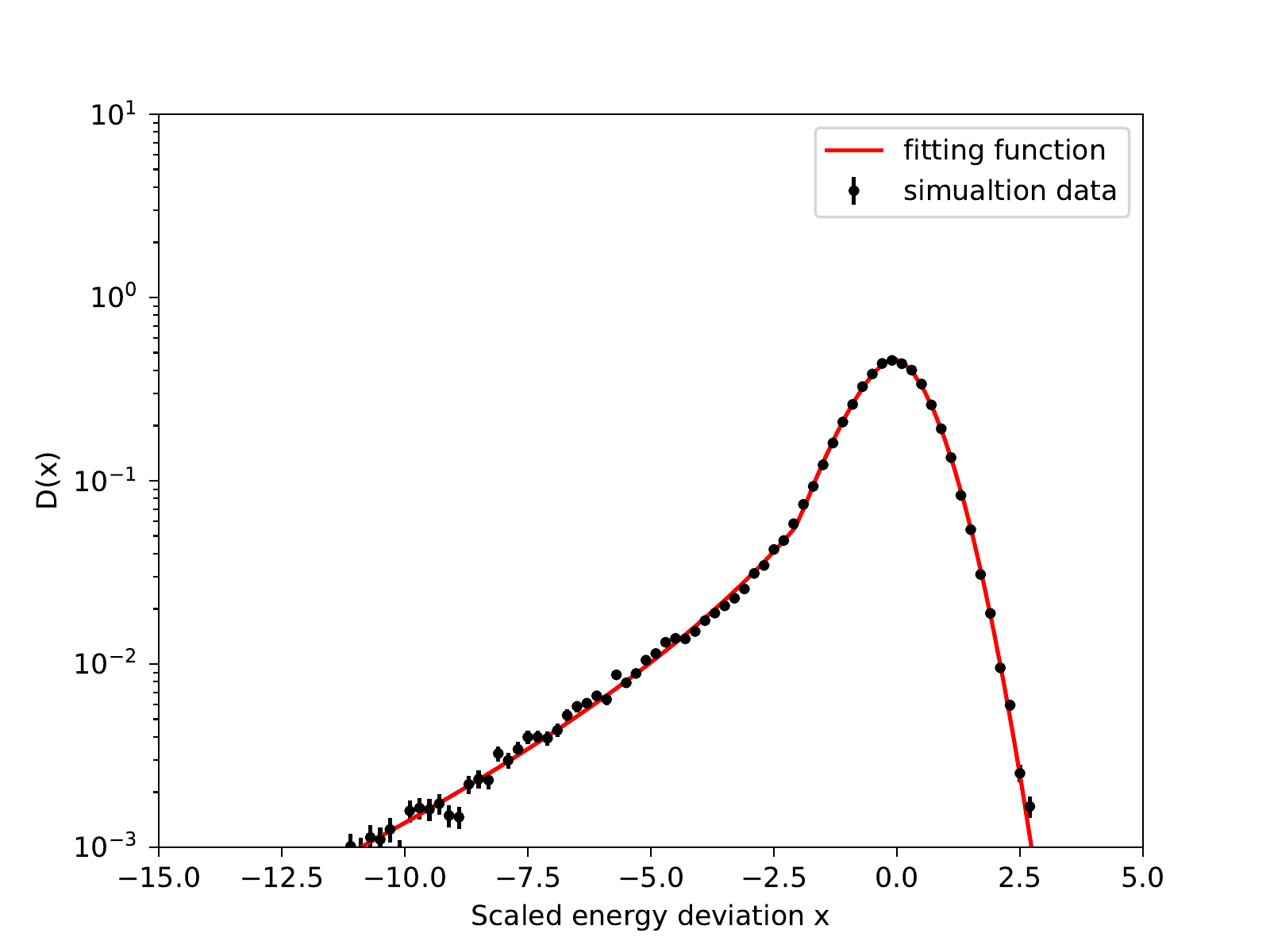}
\caption{The best fit to the scaled energy deviation with the energy dispersion function
         in the energy range [3.16, 10] GeV and incidence angle range [25.84$^{\circ}$, 36.87$^{\circ}$] for HET events.
         The points are the scaled deviation distribution of the MC data, and the line is the best fit function.
         The reduce $\chi^{2}$ od this fitting is 1.05.}
\label{edisp}
\end{figure}

We divided the MC data with the same binned method with the PSF.
Fig. \ref{EDISP} shows the energy resolution of DAMPE for gamma-ray observation at different energy and incidence direction.
And fit the energy dispersion with above function in each bin.
Fig. \ref{edisp} shows an example of energy dispersion fitted with the function
in the bin of $E\in[3.16, 10]\,{\rm GeV}$ and $\theta\in[25.84^{\circ}, 36.87^{\circ}]$ for HET photons.

\section{Observing time and Exposure}

\begin{figure}
\center
\includegraphics[width=10.0cm]{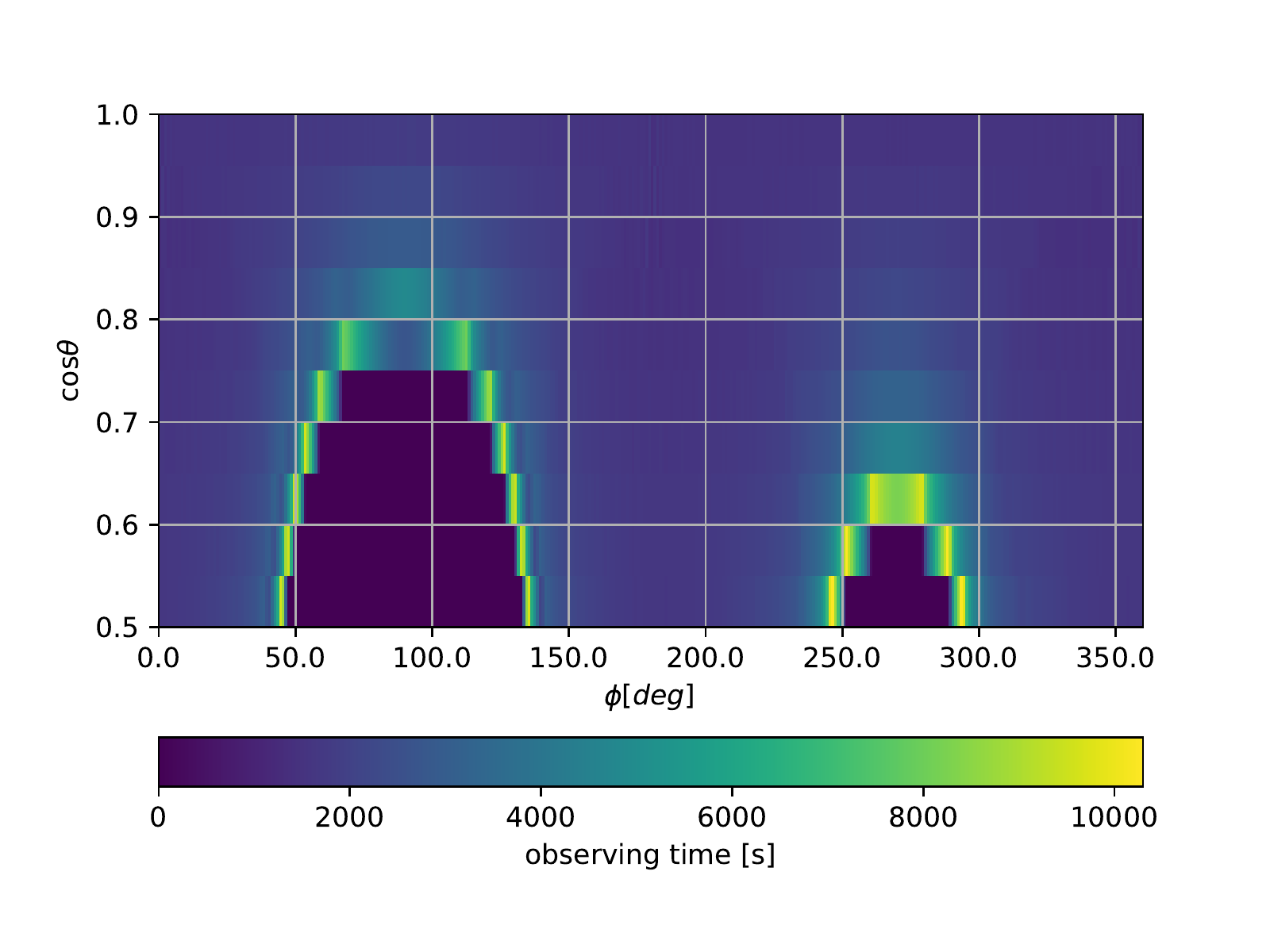}
\caption{The observing time map in the detector reference frame for DAMPE pointing to the Vela pulsar in the first operation year.}
\label{obstime}
\end{figure}

For a particular source in the sky, its direction in the detector reference frame varies with the time.
Since the IRFs various appreciably across the DAMPE field of view (FoV), 
we define the exposure $\epsilon$ for any given energy $E$ and direction in the sky $\hat{p}$
as the integral of the effective area over the time range of interest,
\begin{eqnarray}
\epsilon(E, \hat{p}) = \sum_{s} \int A_{\mathrm{eff}}(E, \hat{v}(t, \hat{p}), s) \mathrm{d}t.
\end{eqnarray}

\begin{figure}
\center
\includegraphics[width=10.0cm]{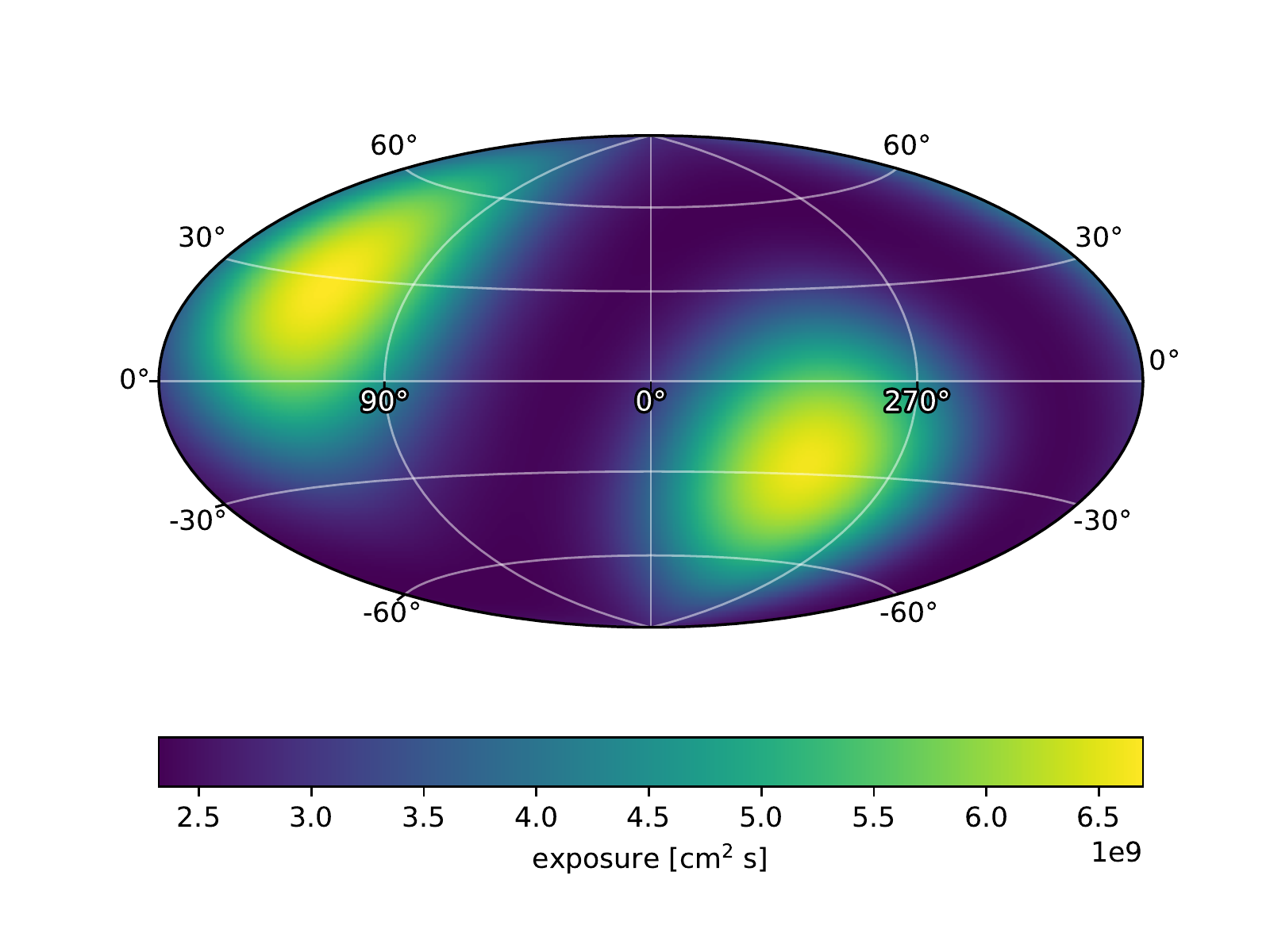}
\caption{The exposure map of DAMPE at 10 GeV in the first year shown in a Hammer-Aitoff projection in Galactic coordinates.
         The maximum value is at the two poles of the equatorial coordinates, while the minimum value is at the equator.}
\label{exposure}
\end{figure}

The exposure can also be expressed as an integral over the solid angle in the detector reference frame,
\begin{eqnarray}
\epsilon(E, \hat{p}) &=& \sum_{s} \int A_{\mathrm{eff}}(E, \hat{v}, s) t_{\mathrm{obs}}(\hat{v}; \hat{p}) \mathrm{d}\Omega \nonumber \\
&=& \int A_{\mathrm{eff}}^{\mathrm{LET}} t_{\mathrm{obs}} \mathrm{d}\Omega
  + \int A_{\mathrm{eff}}^{\mathrm{HET}} t_{\mathrm{obs}} \mathrm{d}\Omega,
\label{eq:exp}
\end{eqnarray}
here the $t_{\mathrm{obs}}(\hat{v}; \hat{p})$ is named {\it observing time} and defined as the total time in the range of interest 
during which DAMPE have observed the direction $\hat{p}$ with detector frame direction $\hat{v}$.
The $A_{\mathrm{eff}}^{\mathrm{LET}}$ and $A_{\mathrm{eff}}^{\mathrm{HET}}$ in Eq. (\ref{eq:exp}) are the effective area for LET and HET photons, respectively.
As an example, we show the observing time map in the detector reference frame for the Vela pulsar in Fig. \ref{obstime}.
With the observing time map and the DAMPE effective area, the exposure then can be calculated according to Eq. (\ref{eq:exp}).
Fig. \ref{exposure} shows the all-sky exposure map of DAMPE at 10 GeV for the first year of operation.
Because DAMPE is in a sun-synchronous orbit, we can see that the exposure is not uniform over the sky.

\renewcommand{\thefootnote}{\arabic{footnote}}
\section{Maximum Likelihood Analysis}
Analyzing the gamma-ray data from DAMPE requires the maximum likelihood method due to the limited number of photons and the angular resolution.
We characterize a source by its photon flux density $F(E, \hat{p}, t; \vec{\lambda})$.
In order to reduce the computational burden, we assume the source is stationary during the time range in each likelihood analysis.
\footnote{For the variable source, the time dependence of the flux can be achieved by repeating the analysis in finer time bins.}
The model of gamma-ray source then can be modeled by
\begin{eqnarray}
F(E, \hat{p}; \vec{\lambda}) = S(E; \vec{\lambda}) M(\hat{p}).
\label{eq:srcmdl}
\end{eqnarray}

Here the $M(\hat{p})$ is a normalized function describing the spatial morphology of the source.
For the point source, the spatial distribution can be described with the Dirac delta function,
$M(\hat{p}) = \delta(\hat{p} - \hat{p}_{0})$, where $\hat{p}_{0}$ is the direction of the point source.
The $S(E; \vec{\lambda})$ in Equation (\ref{eq:srcmdl}) is the spectrum of the source with its parameters $\vec{\lambda}$.

To remove the $\theta$ dependence of the PSF and energy dispersion,
we calculate the exposure-weighted PSF and energy dispersion for any sources included in the analysis:
\begin{eqnarray}
\overline{P}(\delta v; E) =
\frac{\sum_{s} \int P(\delta v; E, \theta, s) A_{\mathrm{eff}}(E, \theta, \phi, s) t_{\mathrm{obs}}(\theta, \phi) \mathrm{d}\Omega}
     {\sum_{s} \int A_{\mathrm{eff}}(E, \theta, \phi, s) t_{\mathrm{obs}}(\theta, \phi) \mathrm{d}\Omega},
\end{eqnarray}

\begin{eqnarray}
\overline{D}(E'; E) =
\frac{\sum_{s} \int D(E'; E, \theta, s) A_{\mathrm{eff}}(E, \theta, \phi, s) t_{\mathrm{obs}}(\theta, \phi) \mathrm{d}\Omega}
     {\sum_{s} \int A_{\mathrm{eff}}(E, \theta, \phi, s) t_{\mathrm{obs}}(\theta, \phi) \mathrm{d}\Omega}.
\end{eqnarray}

Considering the excellent energy resolution of DAMPE 
(i.e., $\sim5\,\%$ at 1 GeV and $\sim1\,\%$ at 100 GeV (\citealt{CHANG20176})),
the influence of energy dispersion can be ignored for most gamma-ray science analysis.
The only exception is the case of searching for narrow-line feature in the gamma-ray spectrum 
(\citealt{fermi15line,liang16gclsLine,LiShangbox}),
which will be performed with other dedicated code.
So in the DmpST we ignore the energy dispersion and regard the measured energy as the true photon energy in current version, and it will be considered in the future if the statistic allows.

With the parameterized source model, the exposure and the exposure-weighted PSF,
we can calculate the model predicted photon rate in the bin $i$ (centered on $E_{i}, \hat{p}'_{i}$) from the source $j$:
\begin{eqnarray}
r_{ij}(E_{i}, \hat{p}'_{i}; \vec{\lambda}_{j}) = \int \mathrm{d}\Omega
F_{ij}(E_{i}, \hat{p}; \vec{\lambda}_{j}) \epsilon(E_{i}, \hat{p}) \bar{P}(\hat{p}'_{i}; \hat{p}, E_{i}).
\end{eqnarray}

The predicted photon rates are compared to the observation data to determine the model parameters.
The information we can get from the DAMPE observation is the energy ($E$),
the direction ($\hat{p}'$) and the time of arrival ($t$) of each photon.
We bin the photons in the region-of-interest (ROI) into a counts cube according to their measured energies and directions.
For each bin, the photon number $N$ follows the Poisson distribution with unknown mean $R$:
$p(N;R) = {R^{N}}/{N!}\cdot\exp(-R)$.
Taking into account all the bins with numbers $\{N_{i}\}$, the Poisson distribution becomes
\begin{eqnarray}
p(\{N_{i}\}; \{R_{i}\}) = \prod_{i=1}^{N_{\mathrm{bins}}} \frac{R_{i}^{N_{i}}}{N_{i}!}\exp(-R_{i}).
\end{eqnarray}

Because of the broad PSF of DAMPE and the strong Galactic diffuse background,
the photons in each bin may originate from multiple sources,
the parameters of which should be determined simultaneously utilizing the likelihood fitting.
With the model predicted photon rates and the real observed data, and based on the Poisson statistics, 
we construct the binned likelihood function (in logarithm form) by summing over all $N_{\mathrm{bins}}$ bins and all $N_{\mathrm{s}}$ sources:
\begin{eqnarray}
\log L(\vec{\lambda}) &=&\sum_{i=1}^{N_{\mathrm{bins}}}
\left(- \sum_{j=1}^{N_{\mathrm{s}}} R_{ij}
      + N_{i} \log \sum_{j=1}^{N_{\mathrm{s}}} R_{ij} \right)\nonumber\\
    &=& \sum_{i=1}^{N_{\mathrm{bins}}}
\left(- \int \mathrm{d}t \int \mathrm{d}E \int \mathrm{d}\Omega' \sum_{j=1}^{N_{\mathrm{s}}} r_{ij}(\vec{\lambda}_{j})
      + N_{i} \log \int \mathrm{d}t \int \mathrm{d}E \int \mathrm{d}\Omega' \sum_{j=1}^{N_{\mathrm{s}}} r_{ij}(\vec{\lambda_{j}}) \right),
\label{eq:like1}
\end{eqnarray}
where the $R_{ij}$ is the model expected photon number in the bin $i$ from source $j$ and the integral is calculated in the corresponding bin $i$ as well.

When the bin widths are taken to be infinitesimal such that only 0 or 1 photon in each bin, 
the summation over $N_{\mathrm{bins}}$ bins becomes to an integral over the whole energy range and the ROI.
Then we get the unbinned form of the likelihood function:
\begin{eqnarray}
\log L(\vec{\lambda}) = - \int \mathrm{d}t \int \mathrm{d}E \int_{\mathrm{ROI}} \mathrm{d}\Omega' \sum_{j=1}^{N_{\mathrm{s}}}r_{j}(\vec{\lambda}_{j})
                          + \sum_{i=1}^{N_{\mathrm{events}}} \log \sum_{j=1}^{N_{\mathrm{s}}}r_{j}(\vec{\lambda}_{j}).
\label{eq:like2}
\end{eqnarray}
By maximizing the likelihood function of (\ref{eq:like1}) or (\ref{eq:like2}), 
we can get the best-fit values of all the free parameters in the source models.

\section{Implementation}

\begin{figure}
\center
\includegraphics[width=10.0cm]{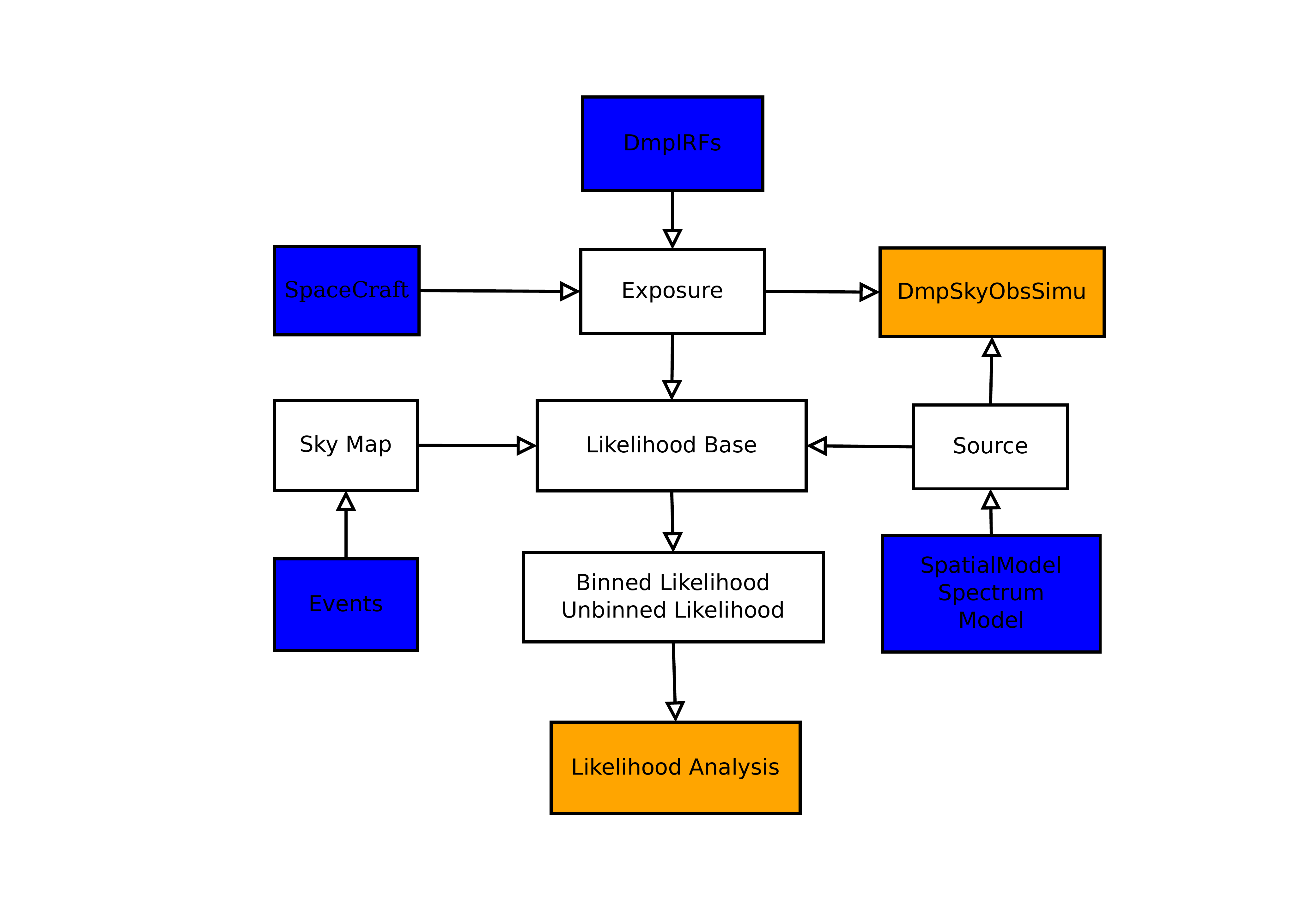}
\caption{The structure of the DmpST. The blue, white and orange represent input, process and output modules, respectively.}
\label{structure}
\end{figure}

The code is coded with Python, based on NumPy (\citealt{numpy}), SciPy \footnote{http://www.scipy.org}, AstroPy (\citealt{astropy})
and iminuit (\citealt{minuit}) packages.
The structure of DmpST is shown in Fig. \ref{structure}.

The input modules are {\tt Events}, {\tt SpaceCraft}, {\tt DmpIRFs}, {\tt Spatial Model}, {\tt Spectrum} and {\tt Model} (shown as blue in Fig. \ref{structure}).
The {\tt Events} module stores the information of photons 
that are selected from all the events detected by DAMPE using the photon selection algorithm (\citealt{Xu2018}).
The information of a photon includes the arrive time ($t$), the reconstructed energy ($E$),
the reconstructed direction in the celestial coordinates ($\alpha_{2000}$, $\delta_{2000}$, $l$, $b$) 
and in the detector reference frame ($\theta$, $\phi$), and the trigger type (s).
The photons of interest in the analysis can be selected according to their times, energies or directions utilizing the {\tt Events} module 
and can be binned into a counts map or a counts cube which is managed by the {\tt Sky Map} module.
The {\tt SpaceCraft} module stores the position, direction and livetime of DAMPE along with time, and can be used to calculate the observing time of DAMPE for any direction in the sky (see Section~3).
The {\tt DmpIRFs} module is used to manage information of instrument response functions (IRFs), 
including the effective area matrix, the parameters of PSF and energy dispersion function.
With these parameters and the fitting functions described in Section 2, 
the distributions of PSF and energy dispersion can be reconstructed.
The {\tt Spatial Model} and {\tt Spectrum} modules provide different kinds of spatial and spectral models of gamma-ray sources, respectively. 
The {\tt Model} module includes all the models of sources those will contribute photons to the ROI.

The process modules comprise {\tt Sky Map}, {\tt Exposure}, {\tt Source}, {\tt Likelihood Base}, {\tt Binned Likelihood} and {\tt Unbinned Likelihood} (white parts in Figure \ref{structure}).
The {\tt Sky Map} module manages the information of counts map or counts cube from the {\tt Events} module,
such as the photon number and celestial coordinates of each bin.
The {\tt Exposure} module calculates the observing time, exposure and the exposure-weighted PSF and energy dispersion
based on the information in {\tt SpaceCraft} and {\tt DmpIRFs} modules.
The {\tt Source} module combines spatial and spectral models based on the {\tt Spatial Model} and {\tt Spactrum} modules for each source in the {\tt Model} module.
The {\tt Likelihood Base} module convolves the PSF with the spatial model, integrals the spectrum over the energy to calculate the expected photons number for each source based on the {\tt Sky Map}, {\tt Exposure} and {\tt Source} modules.
The {\tt Binned/Unbinned Likelihood} modules construct the likelihood function described in Section~4.

Finally, the {\tt Likelihood Analysis} module implements the maximum likelihood estimation with the Minuit algorithm and the basic outputs are the best-fit values ($\hat{\vec{\lambda}}$) of source parameters, the source fluxes and the corresponding statistic uncertainties.
Also we can obtain the confidence level of each source defined as
\begin{eqnarray}
TS_j = -2(\log L(\hat{\vec{\lambda}}_{0,j}) - \log L(\hat{\vec{\lambda}})),
\end{eqnarray}
where $\hat{\vec{\lambda}}_{0,j}$ is the best-fit parameters without source $j$ included in the model.
The $TS_{j}$ follows $\chi^{2}$ distribution with $h - m$ degrees of freedom (\citealt{wilks1938}), 
where $h$ and $m$ are the number of free parameters in the model with/out source $j$.
The {\tt DmpSkyObsSimu} module simulates photons observed by DAMPE with the {\tt DmpIRFs}, {\tt SpaceCraft} and {\tt Source} modules.

\begin{figure}
\center
\includegraphics[width=10.0cm]{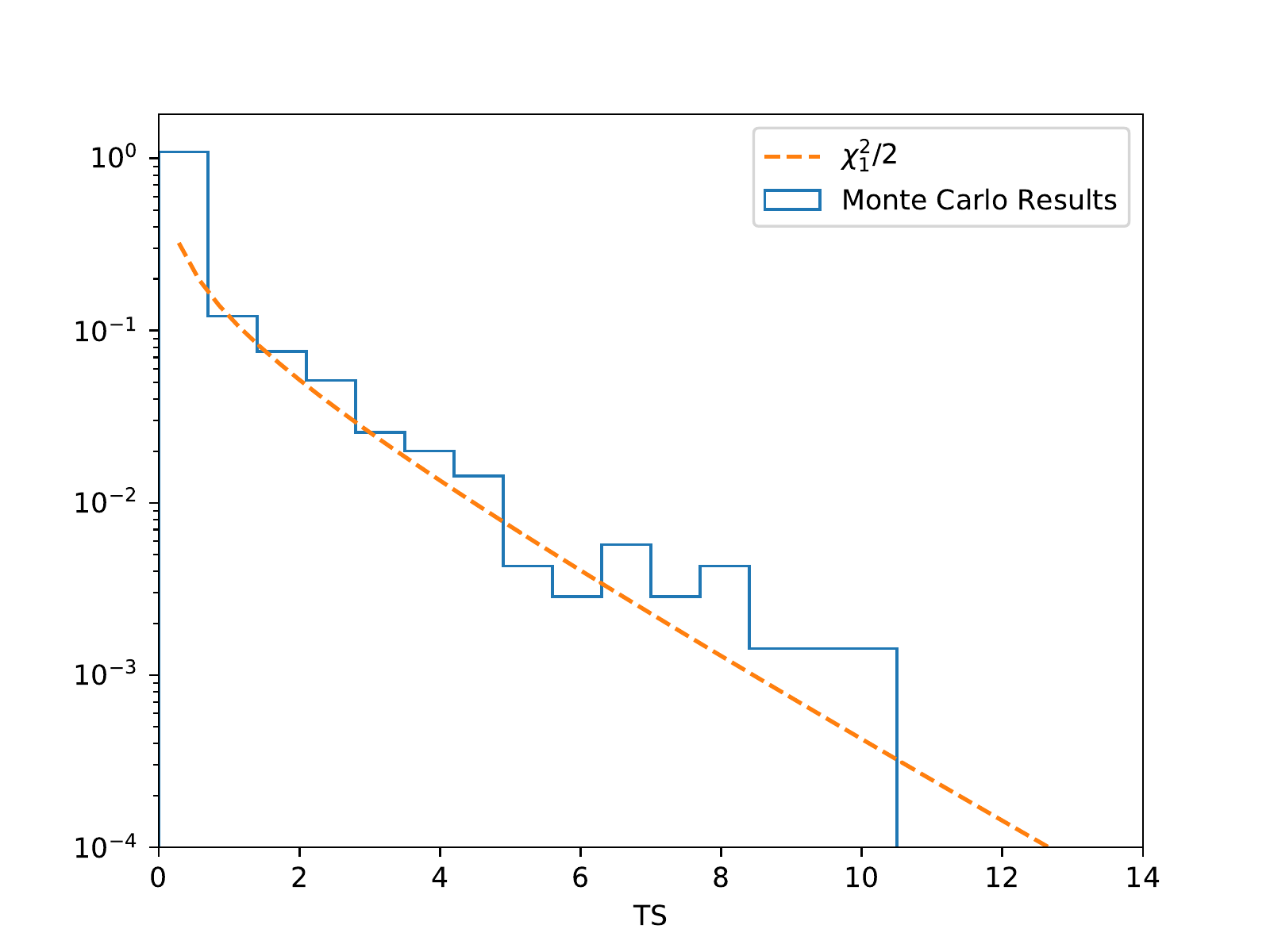}
\caption{The histogram is the normalized distribution of $TS$ values analyzed from simulated data,
and the dash line is the distribution following $\chi^{2}_{1}/2$.
In the analysis, the null hypothesis is no point source and the alternative hypothesis is converse.}
\label{TsValue}
\end{figure}

Monte Carlo simulation has been done with the {\tt DmpSkyObsSimu} module with the Galactic diffuse emission and isotropic emission.
With the {\tt Likelihood Analysis} module, we analyze the simulated data to confirm the distribution of the $TS$.
The null hypothesis is there is no point source, only the background including the Galactic diffuse emission and isotropic emission.
The alternative hypothesis is the converse: there is a point source with Power-Law spectrum with free normalization parameter.
For most point source analysis of DAMPE, the radii of ROI is $\approx 2 \times S_{\mathrm{p}}$ and the typical number of photons $N$ in the ROI is about 25.
Fig \ref{TsValue} shows that for $TS > 0$, the distribution of $TS$ is following $\chi^{2}_{1} / 2$, and the one-half of the simulations have TS = 0 (\citealt{1996ApJ...461..396M}).

\section{Summary}
\label{Summary}

The GeV gamma-ray sky is an important observation target of DAMPE.
To facilitate analyzing the DAMPE gamma-ray data, we have developed a dedicated software named DmpST, 
which implements maximum likelihood analysis to extract the parameters of sources that attribute to the observed gamma-rays.
The DAMPE IRFs that are essential to the gamma-ray data analysis, 
including the effective area, the PSF and the energy dispersion,
are also derived based on high-statistics simulation data.
Making use of the DmpIRFs and DmpST that are detailed in this paper, 
scientific analyses of the gamma-ray data could be carried out to
obtain the best-fit spectral parameters, fluxes and corresponding statistic uncertainties,
and further the spectral energy distribution and light curve of the gamma-ray sources, 
promoting our understanding the nature of high energy gamma-ray phenomena.

\normalem
\begin{acknowledgements}
This work is supported in part by National Key Program for Research and Development (No. 2016YFA0400200),
the Strategic Priority Research Program of Chinese Academy of Sciences (No. XDB23040000),
the 13th Five-year Informatization Plan of Chinese Academy of Sciences (No. XXH13506),
the National Natural Science Foundation of China (Nos. U1631111, U1738123, U1738136, U1738210),
Youth Innovation Promotion Association of Chinese Academy of Sciences,
and the Young Elite Scientists Sponsorship Program.
In Europe the activities and the data analysis are supported by 
the Swiss National Science Foundation (SNSF), Switzerland;
the National Institute for Nuclear Physics (INFN), Italy.
\end{acknowledgements}

\bibliographystyle{raa}
\bibliography{DmpST}

\end{document}